# Some Properties of $B_c$ from Lattice QCD

Seyong Kim

*Center for Theoretical Physics,*

*Seoul National University, Seoul, Korea*

## ABSTRACT

We discuss $B_c$ mass (1S state) and decay constant $f_{B_c}$ calculated by lattice non-relativistic quantum chromodynamcs(NRQCD) method. In leading order of $v^2$, we found that $M_{B_c} = 6.33(2)$ GeV and $f_{B_c} = 395(2)$ MeV where the error bar is the statistical error only. Using these values, we estimate QCD effects to leptonic decay width of $B_c$. The decay width is given by $\Gamma(B_c \to l^+\nu_l) = 0.86(15)m_l^2(1 - 0.0250(2)m_l^2)^2 \times 10^{-14}$(GeV), where $m_l$ is in GeV. Relativistic correction and $\alpha_s$ correction to $f_{B_c}$ have also been considered.

Theoretical understanding on hadronic and electromagnetic annihilation decay of heavy quarkonium system gained a solid footing due to a new factorization theorem by Bodwin, Braaten, and Lepage [1](BBL). They observed that in heavy quarkonium decay processes, typical short distance scale, $M^{-1}$ (Compton wavelength of heavy quark), is well separated from long distance scale, $(Mv)^{-1}$ (size of heavy meson) due to small quark velocity $v$. Then using velocity scaling law of various operators in non-relativistic QCD[2], they showed that decay rates can be written as a sum of factorized products of perturbative coefficients and non-perturbative matrix elements. Compared to earlier calculations, their argument is more consistent because they showed how to do systematic expansion in $v^2$ and $\alpha_s$ which makes the factorization work in every order. It is valid even for P-wave case (or for higher orbital states), where explicit perturbative calculations questioned naive expectation by showing an infra-red divergence in $\mathcal{O}(\alpha_s^3)$[3].

In this work, we are interested in various matrix elements involved in BBL factorization theorem. The matrix elements in this factorization contain non-perturbative information. Such information may be obtained from either fitting experimental data or doing lattice QCD simulation. When there is no available experimental data for the matrix elements, predictions by lattice QCD simulation will be useful. Weak annihilation decay of charmed B meson, $B_c$, is one of such heavy quarkonium systems to which the new factorization can be applied and on which has no experimental data yet. Although there exist studies of $B_c$ system based on models [4, 5, 6], we think that the result from lattice calculation is worthwhile because lattice calculation is based on first principles of field theory so that the result can be improved systematically in terms of accuracy if desired. Interestingly, there are expectations that $B_c$ is within experimental reach of LEP or Tevatron and most certainly of LHC. It is anticipated that decays of $B_c$ will reveal rich information on the Standard model because total annihilation decay of $B_c$ can proceed only through the weak interaction [7, 8, 9].

We have been calculating the matrix elements associated with heavy quarkonium decay such as $c\bar{c}$ and $b\bar{b}$ system using lattice formulation[10]. Extension of our calculational scheme to charmed B meson system is straightforward. Let us describe our method briefly in the following (further details on our lattice simulation method can be found elsewhere [12]). We use non-relativistic lattice QCD (NRQCD) formulation in which the scale separation for the new factorization theorem is most transparent. NRQCD Lagrangian upto $v^2$,

$$\mathcal{L} = \psi^\dagger(D_t - \frac{\vec{D}^2}{2M})\psi + \chi^\dagger(D_t + \frac{\vec{D}^2}{2M})\chi + \mathcal{L}_{light}, \tag{1}$$

is employed in our calculation. Thus our result will have $\mathcal{O}(v^4)$ corrections, which may be as large as 20 %. Also, we cannot address fine structure of the spectrum since such effects require terms of $\mathcal{O}(v^4)$. Here, $\mathcal{L}_{light}$ means just gluon degrees of freedom since we work in the quenched approximation which neglects light quark vacuum polarization effects. We use 200 gauge field configurations on $8^3 \times 32$ lattice volume at $\beta = 5.7$ which were generated by use of Metropolis and over-relaxation algorithm. Although the spatial volume of our lattice configuration is small ($\sim 1.6$ fm), we think that the finite volume effect in our calculation for $B_c$ meson system will be small because the spatial size of $B_c$ itself will be small. We fixed the gauge field to Coulomb gauge prior to the propagator and the matrix element calculation. Under these quenched gauge field background, the quark propagator in lattice NRQCD is calculated as the following [2],

$$G(\vec{x},t+1)_q = (1-H_0/2n)^n U^\dagger(\vec{x},t)(1-H_0/2n)^n G(\vec{x},t)_q + \delta_{\vec{x},\vec{0}}\delta_{t+1,0} \tag{2}$$

with $G(\vec{x},t)_q = 0$ for $t < 0$,

where

$$H_0 = -\Delta^2/2M_0 - 3(1-u_0)/M_0. \tag{3}$$

$\Delta^2$ is the lattice covariant discrete laplacian. The second term ($\equiv E_{sub}$) is the energy shift that arises from the deviation of the plaquette from 1 in mean field theory in "tadpole" improvement scheme [11] and $u_0$ is defined by

$$u_0 = \langle 0|\frac{1}{3}\mathrm{Tr}U_{plaq}|0\rangle^{\frac{1}{4}}, \tag{4}$$

where $U_{plaq}$ is the plaquette value. We choose $n = 2$, $E_{sub} = 0.605524909$ for c quark, and $E_{sub} = 0.154739082$ for b quark and $u_0 = 0.8607297$. Bare c quark mass parameter is chosen to be 0.69, and that of b quark mass is 2.7 in lattice unit. These quark masses are used because they reproduce $J/\Psi$ mass and $\Upsilon$ mass at $\beta = 5.7$. Actually, we need to adjust two quark masses so that the correct $B_c$ meson mass is reproduced. However, since there is no experimental data yet, we use these quark masses. Then, $B_c$ propagator is given by,

$$G(t)_{B_c} = \sum_{\vec{x}} G(\vec{x},t)_q G^{\dagger}(\vec{x},t)_{q'}. \tag{5}$$

Fig 1 is the behavior of $B_c$ propagator. We fit $G(t)_{B_c}$ to the following form,

$$G(t)_{B_c} = Ae^{-Et}. \tag{6}$$

CERN Library MINUIT is used for minimization of the correlated $\chi^2$. We found that the error bar from the single elimination jackknife method is similar to that from the $\chi^2$ method in magnitude. Fig 2 shows effective mass plot. Each points is logarithm of the ratio $G(t)/G(t+1)$ and the error bar is from the jacknife method. The plateau value is 0.955(1) in lattice unit. $B_c$ mass is given by,

$$M_{b\bar{c},c\bar{b}} = \frac{1}{2}(M_{b\bar{b}} - E_{b\bar{b}}a_{b\bar{b}}^{-1} + M_{c\bar{c}} - E_{c\bar{c}}a_{c\bar{c}}^{-1}) + E_{b\bar{c},c\bar{b}}a_{b\bar{c},c\bar{b}}^{-1}, \tag{7}$$

where

$$M_{b\bar{b}} = \frac{3}{4}M_{\Upsilon} + \frac{1}{4}M_{\eta_b}, \tag{8}$$

$$M_{c\bar{c}} = \frac{3}{4}M_{J/\psi} + \frac{1}{4}M_{\eta_c}. \tag{9}$$

$\eta_b$ has not been seen but there exists a good estimate[13]. From PDG [14] and NRQCD group estimate, $M_{c\bar{c}} = 3.06739(56)$ GeV and $M_{b\bar{b}} = 9.45299(142)$ GeV. From our calculation[12], $E_{b\bar{b}} = 0.7948(5)$, $E_{c\bar{c}} = 1.014(2)$ in lattice unit. Here, $a_{b\bar{b}}^{-1}( = 1.37\text{GeV})$ is the lattice spacing for bottomonium, and $a_{c\bar{c}}^{-1}$ (= 1.23GeV) is that for charmonium. We estimate the lattice spacing for $B_c$, $a_{b\bar{c}}^{-1}$ or $a_{c\bar{b}}^{-1}$, by $1.30 = \frac{1}{2}(1.37 + 1.23)$ or $1.30 = \sqrt{1.37 \times 1.23}$ in GeV. This ambiguity in scale is just due to the quenched approximation which can be remedied in the future simulation. The introduced error by this estimate of the lattice spacing is more or less the same in magnitude as the errors resulted from $\mathcal{O}(v^4)$ terms which is not included in our calculation. Thus, by adopting $a_{b\bar{c}}^{-1}$ or $a_{c\bar{b}}^{-1} = 1.30$ GeV as the scale in our results and by using the plateau value $E_{b\bar{c},c\bar{b}} = 0.955(1)$, we get $M_{B_c} = 6.33(2)$(GeV).

The decay constant, $f_{B_c}$, is defined by

$$f_{B_c} p^\mu = \langle 0|\bar{b}\gamma^\mu\gamma_5 c|B_c(p)\rangle. \tag{10}$$

In NRQCD,

$$M_{B_c} f_{B_c} \simeq C_0 \langle 0|\chi^\dagger{}_b \psi_c|B_c\rangle + C_2 \langle 0|(\vec{D}\chi_b)^\dagger \cdot \vec{D}\psi_c|B_c\rangle, \tag{11}$$

with the normalization, $\langle B_c(p)|B_c(p')\rangle = (2\pi)^3 2p_0 \delta^3(p-p')$. The second term is the relativistic correction. The coefficients $C_0$ was calculated upto $\mathcal{O}(\alpha_s)$ in [9]. It is

$$C_0 = 1 + \frac{\alpha_s}{\pi}[\frac{M_b - M_c}{M_b + M_c} \log \frac{M_b}{M_c} - 2]. \tag{12}$$

Also

$$C_2 = -\frac{(M_b + M_c)^2}{8(M_b M_c)^2}. \tag{13}$$

With $M_b^{\rm physical} = 4.5$(GeV), $M_c^{\rm physical} = 1.5$(GeV) and $\alpha_s(\frac{M_b+M_c}{2M_bM_c}) = 0.34$, $C_0 = 1.0(= 0.85)$

without $\alpha_s$ correction (with $\alpha_s$ correction) and $C_2 = -0.09668(\mathrm{GeV}^{-2})$.

Therefore, the matrix elements which is needed for $B_c$ decay rates are,

$$G_1 = \langle 0|\chi_b^\dagger \psi_c|B_c(^1S)\rangle = \sqrt{2M_{B_c}}\widehat{G_1}. \tag{14}$$

$$F_1 = \langle 0|(\vec{D}\chi_b)^\dagger \cdot \vec{D}\psi_c|B_c(^1S)\rangle = \sqrt{2M_{B_c}}\widehat{F_1}. \tag{15}$$

The non-covariant form of $F_1$ is,

$$F_1' = \langle 0|(\vec{\nabla}\chi_b)^\dagger \cdot \vec{\nabla}\psi_c|B_c(^1S)\rangle = \sqrt{2M_{B_c}}\widehat{F_1'}. \tag{16}$$

After removing the exponential fall-off from the binding energy, the asymptotic value of $B_c$ propagator gives the wave function at the origin, $\frac{3\pi}{4}|R(0)|^2 = 0.1125(6)a^{-3}$. From the definition $\widehat{G_1}^2 = \frac{3\pi}{2}|R(0)|^2$, this translates into $\frac{\widehat{G_1}^2}{2} = 0.1125(6)a^{-3}$. Fig 3 shows the ratio of $\frac{2\widehat{F_1'}\widehat{G_1}}{2}$ to the $B_c$ propagator, where the plateau value gives $2\widehat{F_1'}/\widehat{G_1} = 1.813(2)a^{-2}$. On the other hand, we get $2\widehat{F_1}/\widehat{G_1} = 2.400(2)a^{-2}$.

In leading order of $\alpha_s$ (or $v^2$), consistency tells us that $f_{B_c} = 395(2)(\mathrm{GeV})$ from $f_{B_c} = \frac{G_1}{M_{B_c}}$. Beyond leading order, to be consistent in perturbation expansion, $\alpha_s$ correction in $C_0$ should be considered, and the second matrix element (relativistic correction term) need to be included and the lagrangian should contain $\mathcal{O}(v^4)$ terms because $\alpha_s \sim v^2$[1]. In addition, $\overline{MS}$ matrix elements need to be extracted from the calculated lattice matrix elements since the perturbative coefficients, $C_0$ and $C_1$, in decay constant is calculated in $\overline{MS}$ scheme [9]. This requires operator matching between these two different regularization scheme. Such operator matching is similar to $b\bar{b}$ system[15] except the fact that the mass of quark is different from that of anti-quark. For $G_1^{\overline{MS}}$, the feynman diagram, Fig 4a, needed to be evaluated with zero external momentum and the scale of the diagram needed to be set. The algebraic expression for the diagram in coulomb gauge is given by

$$g = \frac{1}{8} g^2 C_F \int \frac{d^3 q}{(2\pi)^3} \frac{1}{\sum_i \sin^2 \frac{q_i}{2}} \left( \frac{1}{1 - f_1^2 f_2^2} - \frac{1}{1 + f_1^2 f_2^2} \right), \tag{17}$$

where $f_i = 1 - \frac{1}{2M_i} \sin^2 \frac{q_i}{2}$($M_i$ is either charm quark (0.8016) or bottom quark (3.137) mass in lattice unit). The numerical evaluation of the integral gives $-0.128298(4)\alpha_s$. By use of $\alpha_s = 0.188(p = \pi/a)$ at $\beta = 5.7$, the integral is equal to $-0.0241220(7)$ numerically. Since the modification due to matching is small, we do not do elaborate scale fixing for the diagrams. If we take the modification due to matching and $\alpha_s$ corrected coefficient $C_0$ into consideration, we get $f_{B_c}^{(1)} = \frac{C_0}{1+g} \frac{G_1}{M_{B_c}} = 344(2)$ MeV from the first term of (11). Since our $G_1$ is $\mathcal{O}(v^2)$ lagrangian result, the above modification is not directly useful in the phenomenological discussion. It just gives us a rough idea on the magnitude of each contributions. On the other hand, similar consideration on our $F_1$ is relevant.

For $F_1^{\overline{MS}}$, we need to subtract a divergent piece which is proportional to $\chi^\dagger \psi(a)$ from feynman diagrams, Fig 4b, 4c, and 4d. Such piece can be deduced from the external momentum independent portion of these diagrams. The continuum $\vec{D} \cdot \vec{D}'$ is the $\vec{p} \cdot \vec{p}'$ dependent portion of these diagrams, which can be obtained by taking zero external momentum limit after taking derivative with respect to external momentum. Similarly, for $F_1^{\overline{MS}}$', a divergent piece from feynman diagrams, Fig 4b and 4c, need to be subtracted and the continuum $\vec{\nabla}\cdot\vec{\nabla}'$ is the $\vec{p}\cdot\vec{p'}$ dependent portion of these diagrams. Thus

$$\widehat{F_1^{\overline{MS}}} = \frac{1}{1+e}(\widehat{F_1^L} - \frac{f}{1+g}\widehat{G_1^L}), \tag{18}$$

where $e$ is $\vec{p}\cdot\vec{p'}$ dependent portion of the diagram 4b and 4c, and $f$ is the subtraction piece. With $\alpha_s(\pi/a) = 0.188$, numerically, $e = (2.6866(1) - 0.35689(7))\alpha_s = 0.43798(9)$ and $f = (2.6743(1) + 5.1913(4))\alpha_s = 1.4787(1)$ for $\widehat{F_1^L}$ and $= 2.6743(1)\alpha_s = 0.50277(2)$ for $\widehat{F_1^{L'}}$. Using $\widehat{F_1^{L'}}$, we get $\widehat{F_1^{\overline{MS}}} = 0.324(2)(\mathrm{GeV})^{7/2}$. On the other hand, using $\widehat{F_1^L}$, we get $-0.260(10)(\mathrm{GeV})^{7/2}$. If we use the non-covariant operator result, relativistic correction gives $f_{B_c}^{(2)} = -17.6(1)$MeV from $f_{B_c}^{(2)} = C_2\frac{F_1}{M_{B_c}}$. However, since the difference between the covariant result and the non-covariant result is significant (even the sign is different), extracting $F_1^{\overline{MS}}$ from lattice study needs further study. Thus in the following discussion, we use only leading order result.

By use of the leading order $M_{B_c}$ and $f_{B_c}$, the leptonic decay width [9],

$$\Gamma(B_c \to l^+\nu_l) = \frac{1}{8\pi}|V_{bc}|^2 G_F^2 M_{B_c} f_{B_c}^2 m_l^2 (1 - \frac{m_l^2}{M_{B_c}^2})^2, \tag{19}$$

becomes

$$\Gamma(B_c \to l^+ \nu_l) = 0.86(15) m_l^2 (1 - 0.0250(2) m_l^2)^2 \times 10^{-14} (\mathrm{GeV}), \tag{20}$$

where $V_{bc} = 0.040(5), G_F = 1.16639(2) \times 10^{-5}\mathrm{GeV}^{-2}$ [14] and $m_l$ in GeV have been used. Since the major source of errors in the calculated decay width is $V_{bc}$, we will get better information on $V_{bc}$ once the leptonic decay width is measured experimentally.

In conclusion, using non-relativistic lattice formulation of QCD, we calculated 1S state $B_c$ mass (= 6.33(2)GeV) and the decay constant $f_{B_c}$(= 395(2)MeV). Using these non-perturbative information, the leptonic decay width of $B_c$ has been discussed.

**Acknowledgments**

We would like to thank Shigemi Ohta of the Institute of Physical and Chemical Research (RIKEN) where this work was initiated for his hospitality and would like to thank prof. H.S. Song and prof. C.Lee of the Center for Theoretical Physics at Seoul National University where this work was completed. Computation Center at RIKEN is gratefully acknowledged for allowing access to their DEC Alpha workstation cluster. Special thanks go to D.K.Sinclair for making his gauge field configurations available to us and for numerous discussions. Comments by G.T.Bodwin, C.T.H.Davies, P.Ko, and H.S.Song were helpful in completing this work.

| quantity | our result | Eichten and Quigg | Chang and Chen | Kiselev et al |
|---|---|---|---|---|
| $M_{B_c}$ (GeV) | 6.33(2) | $6.194 \sim 6.292$ | $6.34 \sim 6.35$ | $6.301 \sim 6.344$ |
| $f_{B_c}$ (MeV) | 395(2) | $479 \sim 687$ | $422 \sim 450$ | $456 \sim 510$ |

Table 1: comparison of $M_{B_c}$ and $f_{B_c}$ from various studies

## Figure captions

1. $B_c$ propagator as a function of $t$ in log scale.

2. effective $E_{b\bar{c},c\bar{b}}$ as a function of $t$.

3. ratio of the matrix element $2F_1'$ to $B_c$ propagator as a function of $t$.

4. feynman diagrams necessary for operator matching. Vertex factor for a) and d) are unit operator, and that for b) and c) are lattice $\vec{\nabla}\cdot\vec{\nabla}'$. Dotted line means temporal component of the gauge field and wavy line means spatial component of the gauge field.

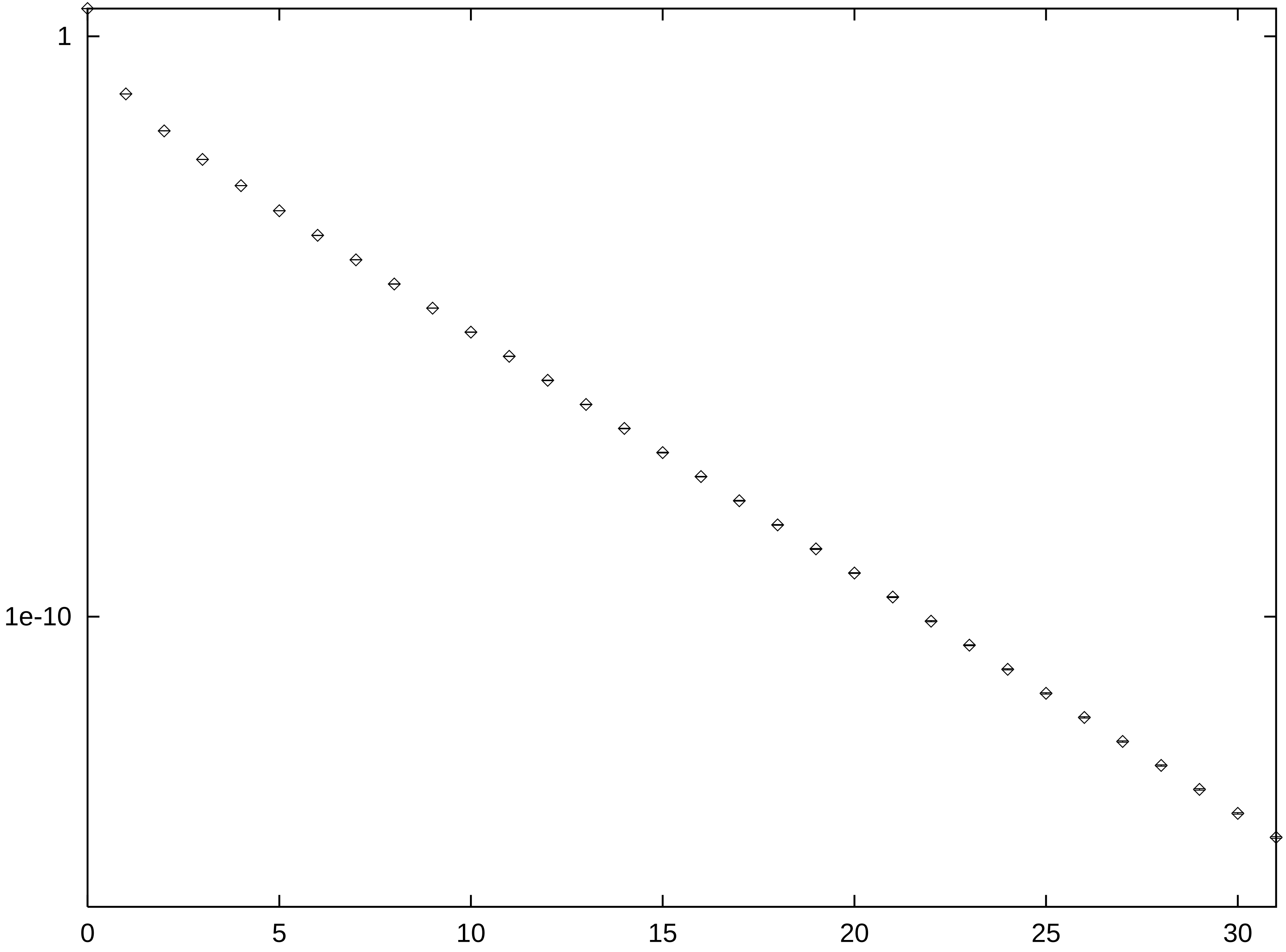

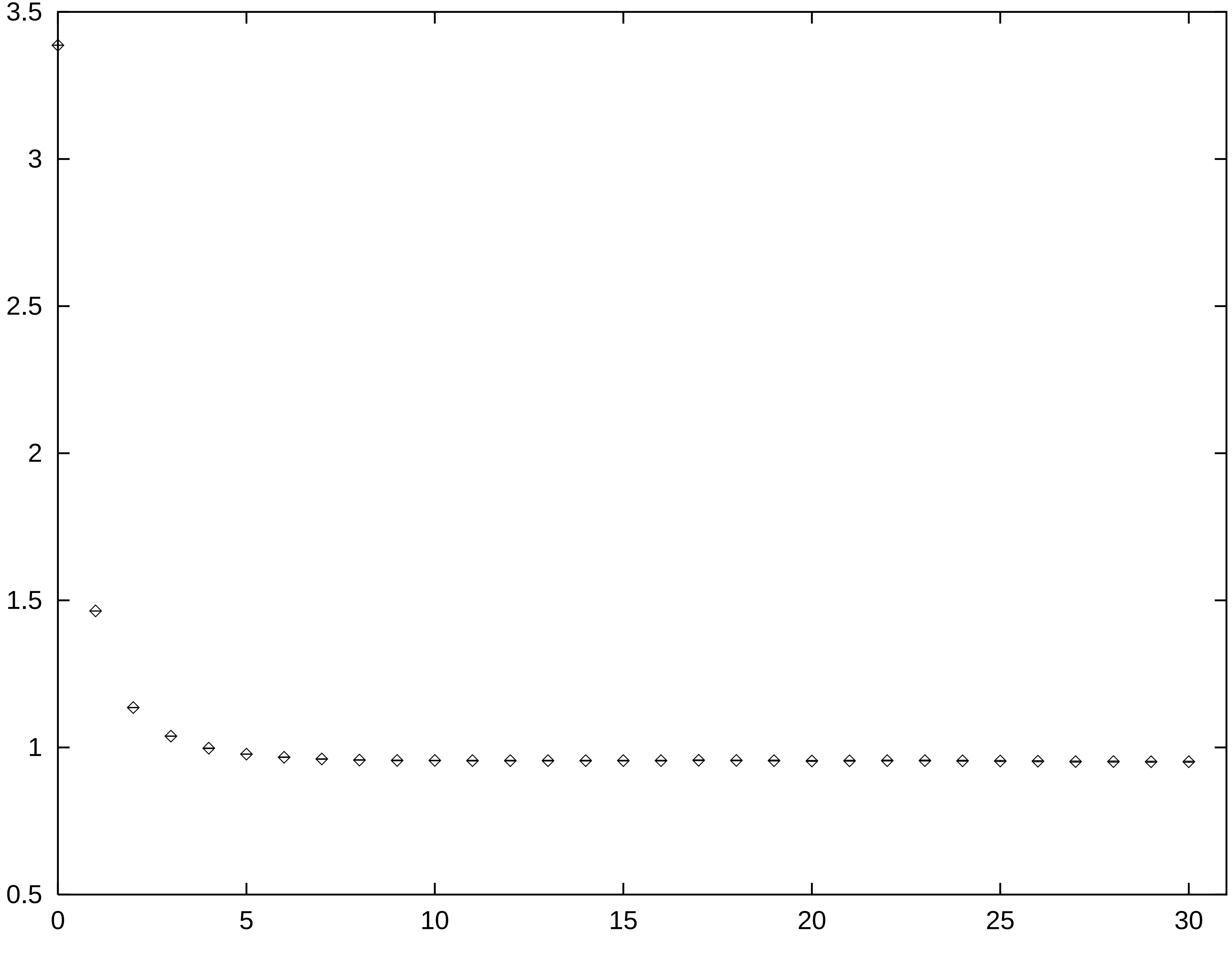

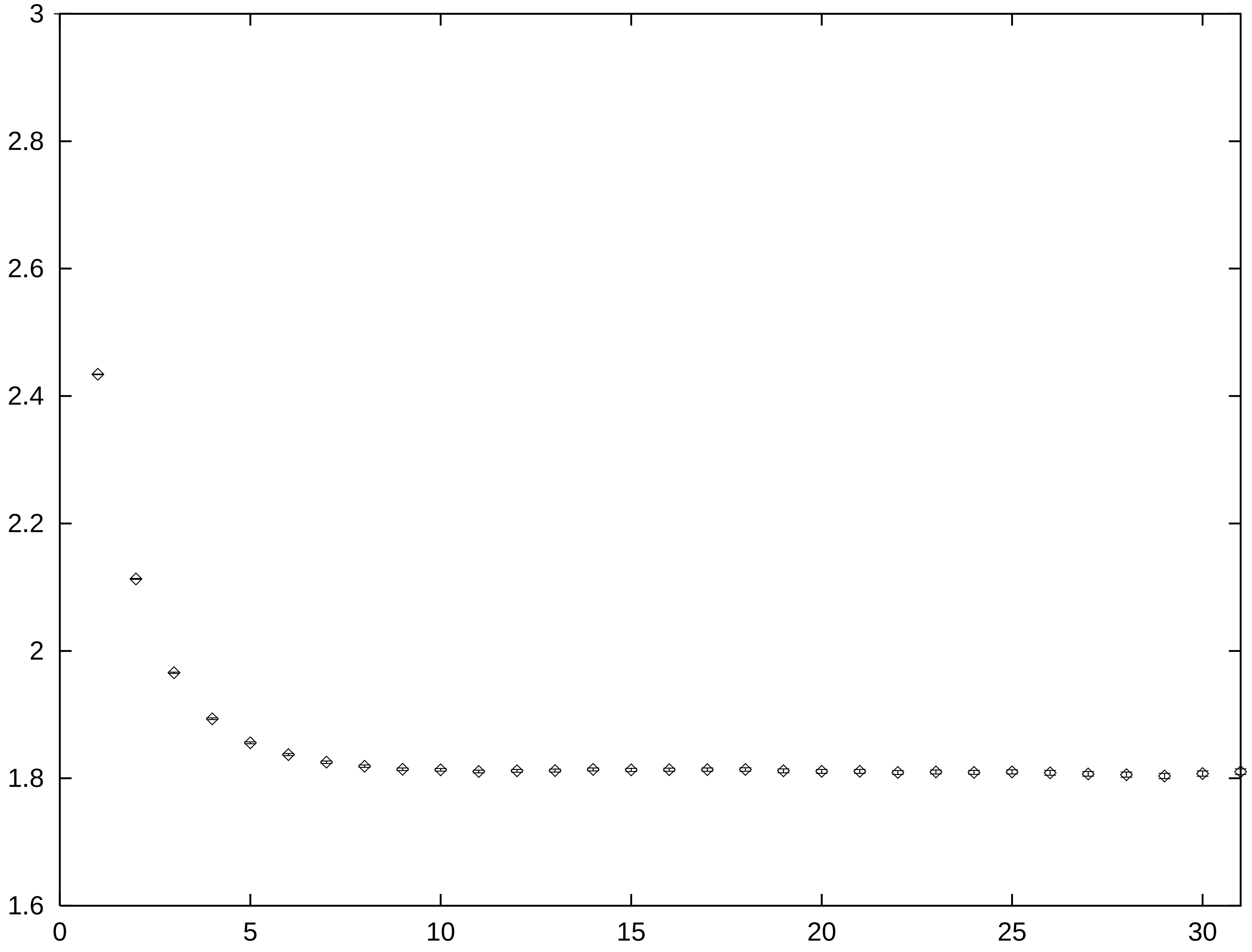

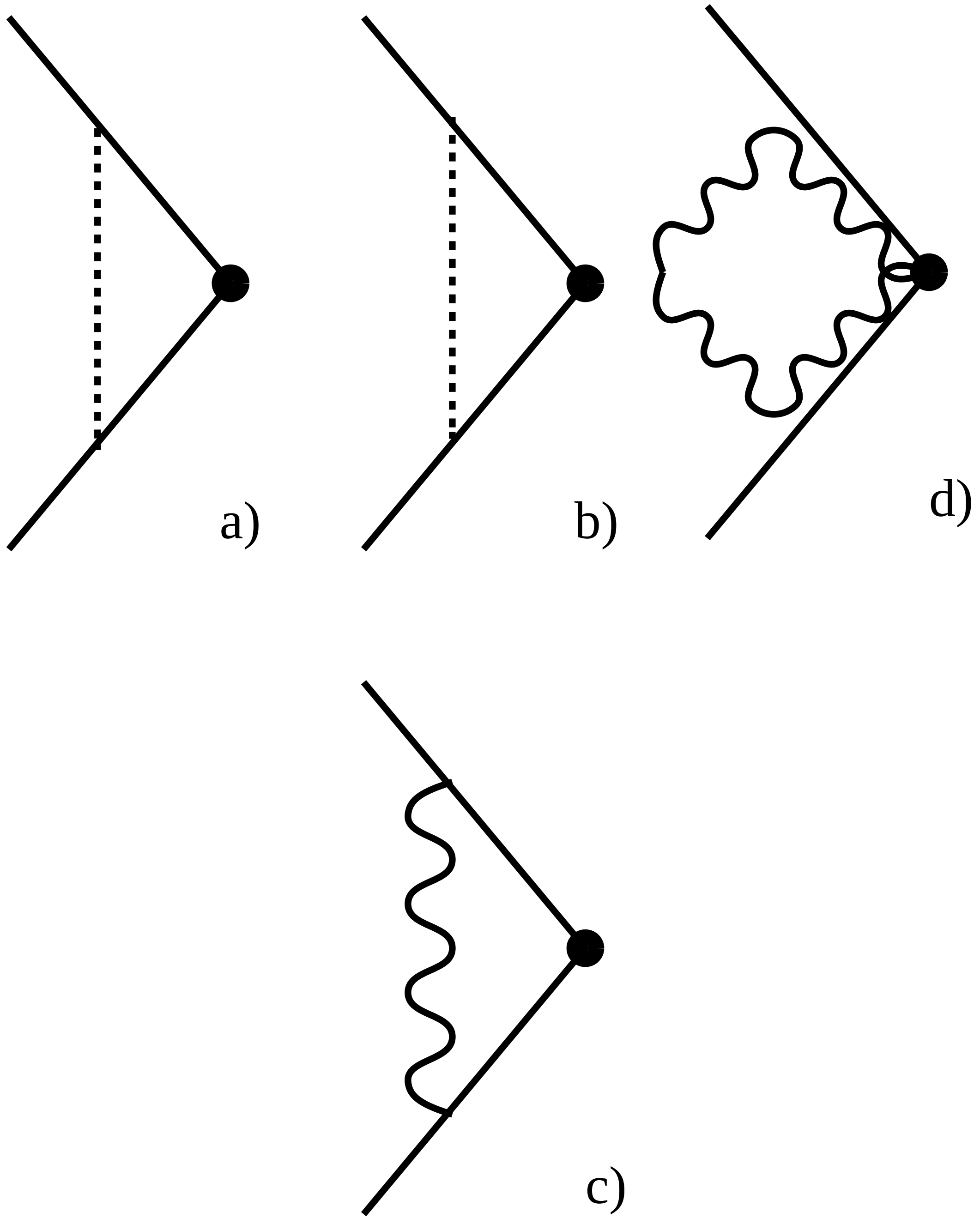
a)
b)
d)
c)